\documentclass[twocolumn]{article}
\usepackage{graphicx} 
\usepackage{amsmath}
\title{A Survey on Large Language Models for Personalized and Explainable Recommendations}
\author{Junyi CHEN}
\date{December 2023}

\begin{document}

\maketitle

\section*{Abstract}
In recent years, Recommender Systems(RS) have witnessed a transformative shift with the advent of Large Language Models(LLMs) in the field of Natural Language Processing(NLP). These models such as OpenAI's GPT-3.5/4, Llama from Meta, have demonstrated unprecedented capabilities in understanding and generating human-like text. This has led to a paradigm shift in the realm of personalized and explainable recommendations, as LLMs offer a versatile toolset for processing vast amounts of textual data to enhance user experiences. To provide a comprehensive understanding of the existing LLM-based recommendation systems, this survey aims to analyze how RS can benefit from LLM-based methodologies. Furthermore, we describe major challenges in Personalized Explanation Generating(PEG) tasks, which are cold-start problems, unfairness and bias problems in RS. 
\section{Introduction}
Recommendation systems are pivotal in aiding users to discover pertinent and personalized items or content. The advent of Large Language Models (LLMs) in Natural Language Processing (NLP) has sparked increased enthusiasm for leveraging the capabilities of these models to elevate and improve recommendation systems.

With the thriving of pre-training in NLP, many language models have been pre-trained on large scale unsupervised corpora, and then fine-tuned for downstream tasks. The transformer architecture\cite{2017attention}, was introduced in 2017, has become a foundation in LLMs. It eschewed the sequential nature of recurrent neural networks (RNNs) in favor of a self-attention mechanism, enabling parallelization and significantly improving efficiency in handling sequential data. Based on Transformer architechture, many Pre-trained Language Models(PLMs) have emerged. GPT series\cite{dale2021gpt}, developed by OpenAI, and BERT\cite{devlin2018bert}, developed by Google, represent two prominent approaches to leveraging transformers for PLMs. The key advantage of incorporating PLMs into recommendation systems lies in their ability to extract high-quality representations of textual features and leverage the extensive external knowledge encoded within them.\cite{2023arXiv230203735L} Different from traditional recommendation systems, the LLM-based models can capture contextual information, comprehending user queries, item descriptions, and other textual data more efficiently.\cite{P5} Based on PLMs, fine-tuning strategy involves training the model on a smaller task-specific dataset. This dataset is typically related to a specific application or domain, such as sentiment analysis, text classification, question answering, or recommender systems. However, fine-tuning large language models on specific downstream tasks paradigm usually needs to fine-tunes all of the parameters in a PLM, which is a computational resource consuming process. Most researchers and companies cannot access as much resouce as OpenAI or Microsoft or Google. As a result, a recently proposed paradigm, prompt learning\cite{liu2023pre}, further unifies the use of PLMs on different tasks
in a simple yet flexible manner. In general, prompting is the process of providing addtional information for a trained model to condition while predicting output labels for a task, for example, providing a piece of text inserted in the input examples. Prompt tuning is different from fine tuning in that it only requires storing a small task-specific prompt for each task instead of making a task specific copy of the entire PLM in fine tuning. The advantage of this paradigm lie in two aspects:(1)It bridges the gap between pre-training and downstream objectives, allowing better utilization of the rich knowledge in pretrained models. This advantage will be multiplied when very little downstream data is available. (2)Only a small set of parameters are needed to tune for prompt engineering, which is more efficient.\cite{surveyllm4rec}

This survey focuses on utilizing LLMs for Personalized Explanation Generating task. I will discuss the background of NLP for text generation in Chapter 2, the challenges existed in explainable recommendations in Chapter 3, the current approaches intending to solve these challenges in Chapter 4, and finally, the conclusion in Chapter 5.
\section{Natural Language Generation}
Natural Language Generation(NLG) stands at the forefront of articial intelligence, representing a transformative field that empowers machines to comprehend and generate human-like text. NLG plays a pivotal role, bridging the gap between machines and humans through the creation of coherent, contextually relevant language. Especially, in the recommendation field, NLG approaches have a great potential to be applied to generating justifications for recommending results, which enhances the presuasiveness and effectiveness of Recommender systems.
\subsection{Methodologies}
The early stages of NLG were marked by rule-based approaches, where explicit linguistic rules were crafted to govern the generation of text. These systems operated on predefined templates and grammatical structures, limiting their flexibility and adaptability to diverse applications. A significant leap came with the advent of neural network architectures, particularly recurrent neural networks (RNNs)\cite{RNN4generation}. RNNs allowed for sequential processing, enabling models to better capture contextual dependencies in language. However, they faced challenges with long-term dependencies, limiting their effectiveness in handling complex language structures. The turning point in NLG arrived with the introduction of the transformer architecture\cite{2017attention}, the PLMs based on transformer architecture harnessed the power of attention mechanisms, enabling parallel processing and capturing intricate contextual relationships across words.
\subsubsection{Transformer Architecture}
In this section, we briefly go through Transformer architecture. Transformer consists of n identical layers, the $l$ th layer encodes the previous layer's output \(S_{l-1}\) into \(S_l \in \text{R}^{|S|\times d}\), where \(\left|s\right|\) is the length of the input token sequence, and $d$ denotes the dimension of token embeddings. Each layer is composed of two sub-layers: Multi-Head Self-Attention(MHSA) and Feed-Forward Network. For self-attention mechanism, the input consists of queries and keys of dimension \(d_k\), and values of dimension \(d_v\). The attention scores can be calculated.
    \begin{equation}
         Attention(Q,K,V)=softmax(\frac{QK^T}{\sqrt{d_k}})V
    \end{equation}
where $Q$,$K$,$V$ are the query, key and value matrices obtained by linear transformations of the input sequence. $d_k$ is the dimension of the key vectors. Multi-Head Attention is then calculated:
    \begin{equation}
        MultiHead(Q,K,V)=Concat(head_i,...,head_h)W_O
    \end{equation}
where $head_i=Attention(QW_i^Q,KW_i^K,VW_i^V)$, the parameter matrices are denoted as: $W_i^Q\in \text{R}^{d_{model}\times d_k}$, $W_i^K\in \text{R}^{d_{model}\times d_k}$,$W_i^V\in \text{R}^{d_{model}\times d_k}$, and $W_O\in \text{R}^{hd_v\times d_{model}}$.
In addition to attention sub-layers, each of the layers in the encoder and decoder contains feed-forward network.The FFN section is a two-layer FFN with the ReLU activation function.
    \begin{equation}
        FFN(x)=max(0,xW_1+b_1)W_2+b_2
    \end{equation}
x is the MHSA's output, $W_1$,$b_1$,$W_2$,$b_2$ are weight parameters.
    \begin{figure}
        \centering
        \includegraphics[scale=0.5]{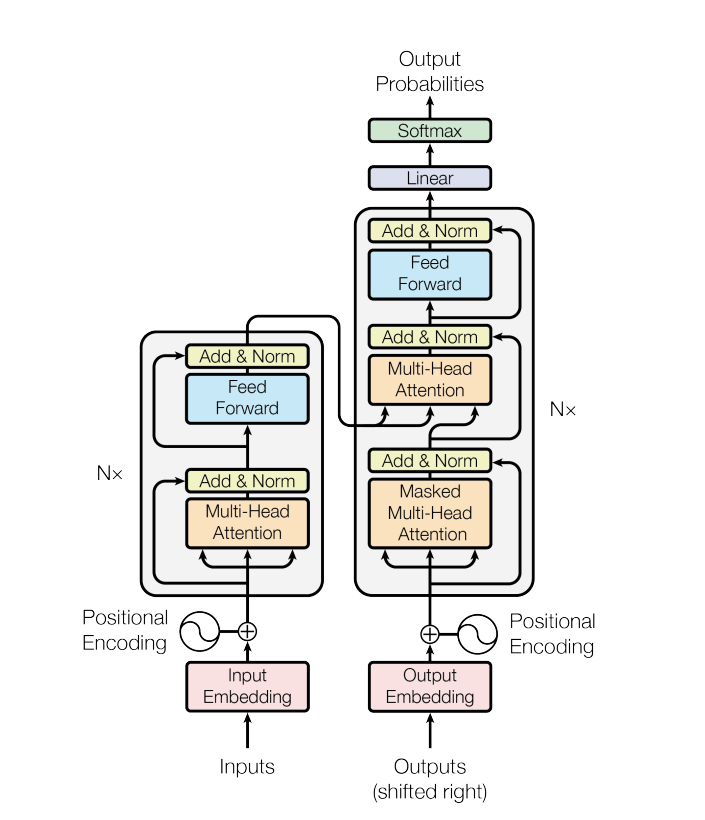}
        \caption{Architecture of Transformer}
    \end{figure}
We focus on Transformer-based pre-training language models. Bidrectional Encoder Representations from Transformer(BERT), Generative Pre-trained Transformer(GPT), and Text-to-Text Transfer Transformer(T5) are three main architectures. BERT is an encoder-only model, it is pretrained by predicting missing words in a bidirectional manner. While GPT is an example of decoder-only model, it predicts the next word in a sequence given its context. T5 is an encoder-decoder model, it treats all NLP tasks as text-to-text problems, where both inputs and outputs are represented as textual sequences. Recent open-source PLM Llama-2\cite{touvron2023llama},developed by Meta, ranges in scale from 7 billion to 70 billion parameters. And its fine-tuned LLMs, called Llama-2-chat, outperform open-source chat models on most benchmarks.
\subsubsection{Training strategies}
Given the significant impact that PLMs have achieved on NLP tasks in the pre-train and fine-tune paradigm, there has been a surge in adapting such paradigms to recommendation tasks. Figure 2 shows a high-level overview of the PLM for recommender systems.\cite{surveyllm4rec}
\begin{figure}
    \centering
    \includegraphics[scale=0.5]{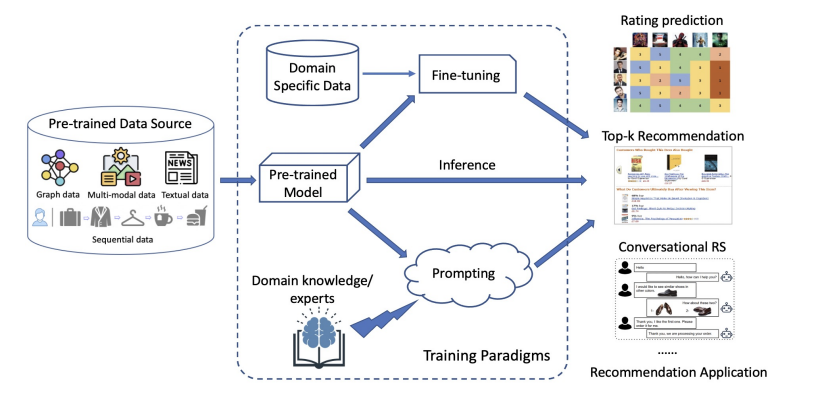}
    \caption{A generic architecture of language modeling paradigm for recommendation purpose}
\end{figure}

Many researchers adopt pre-train strategy. For example, BERT4Rec\cite{sun2019bert4rec} models sequential user behavior with a bidirectional self-attention network through Cloze task, and Transformer4Rec\cite{de2021transformers4rec}adopts a haggingface transformer-based architecture as the base model for next-item prediction. These two models have laid the foundation for LLM-based recommender systems. 

Further more, some not only pre-train the model, but also fine-tune it with different specific datasets. Fine-tuning process will adjust the whole model parameters, for example, Petrov and Macdonald proposed GPTRec\cite{gptrec}, which is a generative sequential recommendation model based on GPT-2. GPTRec is based on generative LLM, while BERT4Rec is based on discriminative LLM. It uses SVD tokenization for memory efficiency. UP5\cite{hua2023up5} is a P5-based model, it includes two sub-modules: a personalized prefix prompt that enhances fairness with respect to individual sensitive attributues and a prompt mixture that integrates multiple counterfactually-fair prompts for a set of sensitive attributes. For parameter-efficient training, they only optimize the parameters in the prefix prompt and leave the PLM for RS untainted. Since tuning the whole model is usually time-consuming and less flexible, many choose to fine-tune only partial parameters of the model to balance between training overhead and recommendation performance.\cite{hou2022towards} This leads to Parameter-Efficient Fine-Tuning(PEFT) approaches. In order to reduce the memory consumption, there are two main lines applied in PEFT. One is to add small neural modules to the PLMs and fine-tune only these modules for each task. Adapter tuning inserts small neural modules between the layers of the base model.  Prompt tuning attaches additional trainable tokens to the input or hidden layers. Another line is considering the incremental updates of the PLMs, without modifying the model architecture. Hu et al.(2022)\cite{2021arXiv210609685H}propose LoRA(\textbf{Lo}w \textbf{R}ank \textbf{A}daptation), which parameterizes the update \(\Delta\) as a low-rank matrix by the product of two small matrices:
\begin{equation}\label{lora}
W=W^{(0)}+\Delta=W^{(0)}+BA 
\end{equation}
During this fine-tuning, where \(\Delta\in\ R^{d_1\times\ d_2}\) ,
we only need to update \(A\in\ R^{r\times d_2}\) and \(B\in\ R^{d_1\times\ r}\). Because the rank \(r\) is much smaller than the dimension of \(W\) , the training overhead can be reduced up to 70\% in total. This technique has significantly improved the implementing of affordable fine-tuning. 

Instead of adapting PLMs to different downstream recommendation tasks by designing specific objective functions for fine-tuning, there is a rising trend to reformulate recommendation tasks through hard/soft prompts. Figure 3 \cite{surveyonllm4rec}shows the detailed explanation of four different training manners for LLM-based recommendations.
\begin{figure}
    \centering
    \includegraphics[scale=0.5]{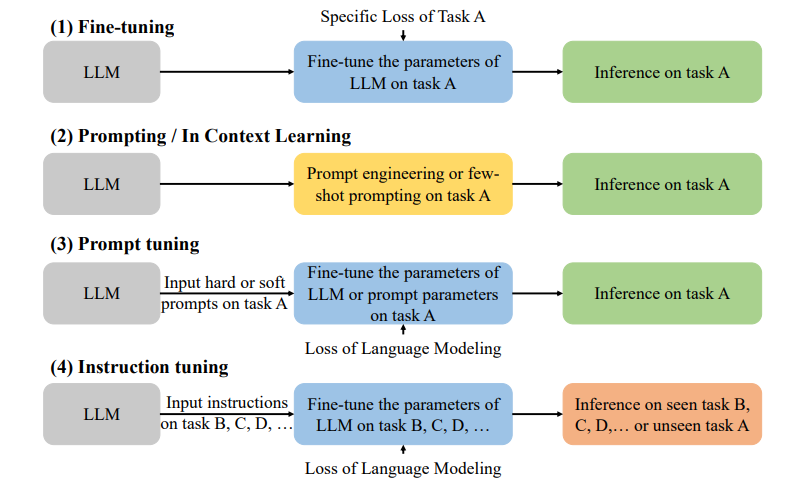}
    \caption{Four different training manners for LLM-based recommendations}
\end{figure}
Generally speaking, prompting paradigm aims to design more suitable instructions and prompts to assist LLMs for better understanding of recommendaiton tasks such as rating predicitons, sequential recommendation, direct recommendation, explanation generation, and review summarization. This prompting paradigm can be categorized into two training strategies, which are Tuning-free prompting and Tuning-required prompting. The former can be refered to as zero-shot recommendations in that this usually generate recommendations or/and subtasks without changing the parameters of PLMs. The latter training strategy either tunes the parameters of PLMs or prompt-relevant parameters, or both of them. For example, Li et al.\cite{LiPEPLER}proposes a continuous prompt learning approach by first fixing the PLM, tuning the prompt to bridge the gap between the continuous prompts and the loaded PLM, and then fine-tuning both the prompt and PLM, resulting in a higher BLUE score. I will discuss this approach in details in the next chapter.
\section{Explanation Generating for Recommendation}
\subsection{PEPLER}
PEPLER\cite{LiPEPLER} is proposed for PEG task. It includes two methods which the author put them as discrete prompt learning and continuous prompt learning. In discrete(hard) prompt learning, a key idea is to represent the IDs in Recommender systems as domain-specific words such as movie titles and item features. This bridges the gap between Recommendation task and Natural language Generation task because IDs are indispensable in Recommender systems. Figure 4 shows a structure of PEPLER-D method, for each user-item pair \((u,i)\), the features in \(F_u\cap F_i\) are more informative because they are both related to user and item. If the size is too small, \((F_u\cup F_i)/(F_u\cap F_i)\) can also be taken into consideration. The input sequence to the pre-trained model is \(S=[f_1,...,f_{|F_{u,i}|},e_1,...,e_{|E_{u,i}|}]\), \(|F_{u,i}|\) and \(|E_{u,i}|\) denote the number of features and explanation words, respectively. we can obtain the sequence final representation \(S_n=[s_{n,1},...,s_{n,|S|}\), a linear layer is applied to each token's final representation to map it onto a \(|\text{V}|\)-sized vector. For example, \(s_{n,t}\) becomes \(c_t\) after passing through this layer:
\begin{equation}
    c_t=softmax(W^vs_{n,t}+b^v)
\end{equation}
The vector \(c_t\) represents the probability distribution over the vocabulary \(\text{V}\), for model learning loss function, Negative Log-likelihood(NLL) is applied.
\begin{equation}
    \mathcal{L}_D=\frac{1}{|\mathcal{T}|} \sum_{(u,i)\in \mathcal{T}} \frac{1}{|E_{u,i}|}\sum_{t=1}^{|E_{u,i}|} -log c^{e_t}_{|F_{u,i}|+t^{\prime}}
\end{equation}
where the probability \(c^{e_t}\) is offset by \(|F_{u,i}|\) positions because the explanation is placed at the end of the sequence.
\begin{figure}
    \centering
    \includegraphics[scale=0.3]{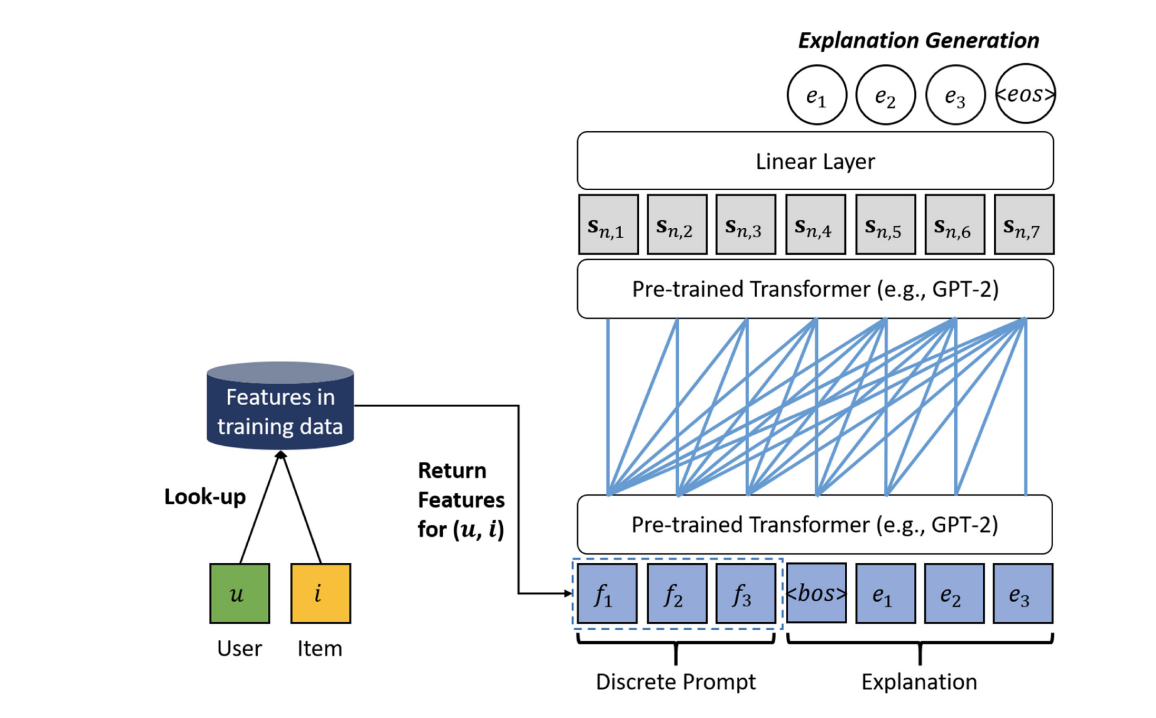}
    \caption{the PEPLER-D method that utilizes item features as a discrete prompt}
\end{figure}
Another method is continuous(or soft) prompt learning. Prompts do not necessarily have to be words or even readable, but can be vector representations. In this case, the input sequence can be represented as\(S=[u,i,e_1,...,e_|E_{u,i}|]\), the user u's vector representation can be retrieved via:\(u=U^Tg(u)\), \(g(u)\in {(0,1)}^{|\mathcal{U}|}\) denotes a one-hot vector, whose non-zero element corresponds to the position that user u's vector locates in \text{U}. The vector \(i\) can be obtained in a similar way. The follow-up steps are identical to discrete prompt learning.
\begin{figure}
    \centering
    \includegraphics[scale=0.3]{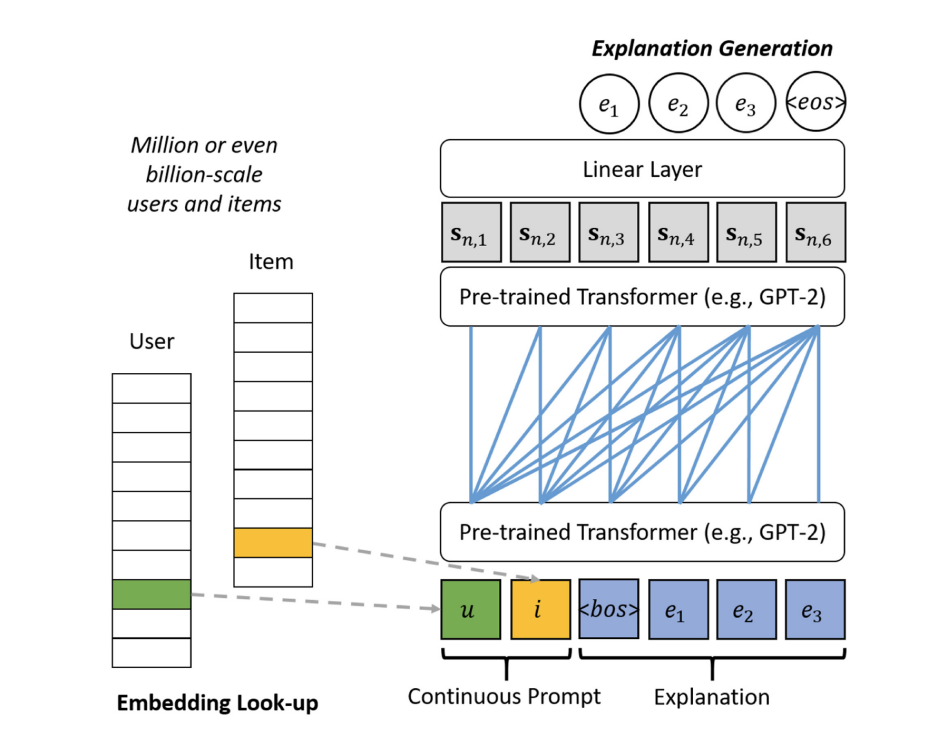}
    \caption{the PEPLER method that treats user and item IDs as continuous prompts}
\end{figure}

\subsection{Rationale Generation and In-text Learning}
Googel research team introduces Chain of Thought(CoT)\textemdash a series of intermediate reasoning steps, which significantly improves the ability of LLM to perform complex reasoning. Providing or automatically generating rationale examples have proven to be effective. However, when tackling different downstream tasks, specific CoT usually need to be designed manually, this is not only time consuming and ineffective, but the perfomances also could vary tremendously. STaR\cite{star_COT} proposes to generate a new rationale by providing the model with the correct answer when each problem failure occurs. These generated rationales are then collected as part of the training data, which improves overall accuracy. Similar approach could be taken into consideration when solving PEG problems.

\subsection{Evaluation Metrics}
Explainability evaluation contains various aspects. For evaluating the quality of the generated explanations, there are different metrics which are widely accepted. BLEU\cite{papineni2002bleu} and ROUGE\cite{lin2004rouge} are two widely used metrics in machine translation and text summarization respectively. BLEU stands for Bilingual evaluation understudy, it calculates the precision by comparing the n-grams(contiguous sequences of n items, like words or characters) in the machine-generated translation with those in the reference translations. ROUGE stands for Recall-Oriented Understudy for Gisting Evaluation, which measures the overlap between the n-grams of the system-generated summary and those reference summaries. 

For evaluating the diversity of the generated explanations, Unique Sentence Ratio(USR), Feature Coverage Ratio(FCR) and Feature Diversity(FD) can be taken into consideration. \cite{metrics1}
\begin{equation}
    USR=|\mathcal{S}|/\mathcal{N}
\end{equation}
where \(\mathcal{S}\) denotes the set of generated unique explanations, and \(\mathcal{N}\) is the number of total explanations.
\begin{equation}
    FCR=\frac{\mathcal{N_g}}{|\mathcal{F}|}
\end{equation}
This metric is to measure how many different features are shown in the generated explanations, \(\mathcal{N_g}\) is the number of distinct features, and \(\mathcal{F}\) is the collection of all features in the dataset.
\begin{equation}
    FD=\frac{2}{N\times (N-1)} \sum_{u,u\prime,i,i\prime}|\hat{\mathcal{F}_{u,i}}\cup \hat{\mathcal{F}_{u\prime,i\prime}}|
\end{equation}
where \(\hat{\mathcal{F}_{u,i}}\) and \(\hat{\mathcal{F}_{u\prime,i\prime}}\) respectively represent two sets of features contained in two generated explanations. This metric is to compute the overall overlap between feature sets, a lower FD scores implies a higher diversity across explanations.

\section{Challenges}
\subsection{Cold-Start Problem}
The cold-start problem includes three cases, new community, new item, and new user\cite{bobadilla2012collaborative}. The new community refers to such recommender systems that lack users' interactions, which make them hard to provide reliable recommendations. The new item refers to the situation that a new item is added to the system, but no interations are peresent. The new user refers to the situation that a recommender system lacks the new user's interaction, which makes it hard to provide personalized recommendations. 

Recently, to solve the cold-start problem, PromptRec\cite{wu2023towards}considers the user and item profile features as input, transforms them into templated sentences, and predicts recommendations through PLMs. Specifically, they first introduce a verbalizer mapping profile features to natural langugage descriptions, the apply a template reformatting the recommendation task as Masked Language Modeling(MLM) task, and finally leverage a PLM to accomplish the task and recommendations. Figure 6 shows the structure of PromptRec. It is also possible that applying a similar scheme to Generating explanations for new users.
\begin{figure*}[t]
    \centering
    \includegraphics[scale=0.4]{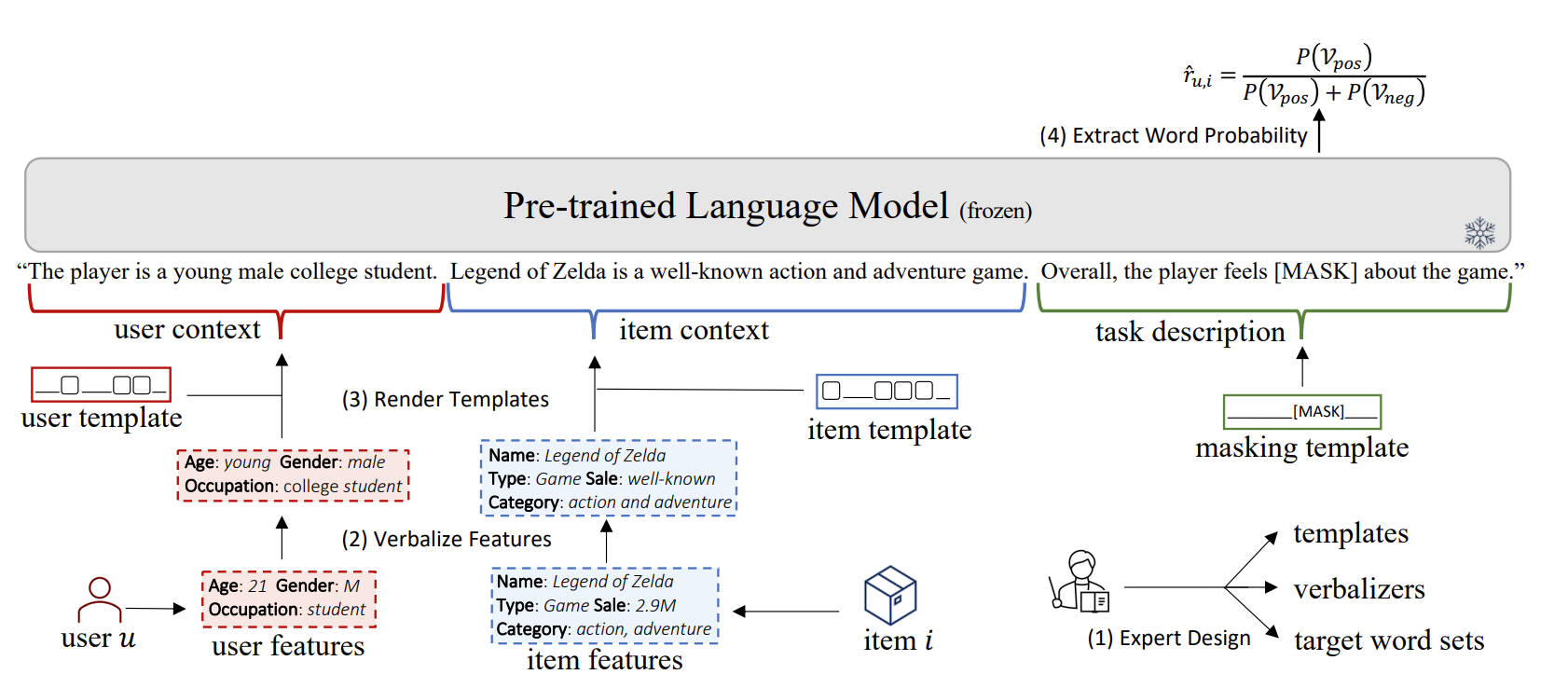}
    \caption{PromptRec:Prompting PLMs to make personalized recommendations under the system cold-start setting}
\end{figure*}
\subsection{Bias Problem}
Bias is common in RS\cite{chen2023bias} for the following components: user behaviors, presented items, feedback loop when training. Firstly, user behavior is observational because user generates behaviors on the basis of the exposed items, making the observational data confouned by the exposure mechanism of the system and self-attention of the user. The presented items are unavoidably biased, some items are more popular than others and thus receive more user behaviors(such as reviews or purchase history). Also, the feedback loop of machine learning can intensify the biases over time. 

One of the most significant bias categories in explanation generating is popularity bias(or long-tail problem), which stands for popular items are recommended more frequently while ignoring other items. In particular, for generated explanations, the unfairness issues will occur, influencing the level of personalized recommendation and the trust of the users eventually. UP5\cite{hua2023up5} intends to solve unfairness problem by designing a personalized prefix prompt that enhances fairness with respect to individual sensitive attributes, which shows an improvement in eliminating unfairness across gender, age, and occupation.
\section{Conclusion}
In conclusion, the integration of Large Language Models (LLMs) has revolutionized Recommender Systems (RS). This paradigm shift holds tremendous potential for personalized and explainable recommendations, as LLMs offer a versatile toolset for processing vast textual data, thereby enhancing user experiences. Despite these advancements, challenges persist in the realm of Personalized Explanation Generating (PEG) tasks, including cold-start problems and issues of unfairness and bias in RS. Addressing these challenges is crucial for optimizing the potential of LLMs in recommendation systems, ensuring not only accuracy but also transparency and fairness in explaining personalized recommendations to users. 
\bibliographystyle{unsrt}
\bibliography{cited}
\end{document}